\documentclass[prx,twocolumn,superscriptaddress,floatfix,nopacs]{revtex4}
\usepackage{graphicx,amsfonts,amssymb,amsmath,hyperref,enumerate}

\newif\ifhyper
\hypertrue
\ifhyper
\hypersetup{
   citecolor = {green},
   colorlinks = {true}, 
   urlcolor = {blue} 
}
\fi

\newcommand{\beq}{\begin{equation}}
\newcommand{\eeq}{\end{equation}}
\newcommand{\beqa}{\begin{eqnarray}}
\newcommand{\eeqa}{\end{eqnarray}}
\newcommand{\ket} [1] {\vert #1 \rangle}

\def\ket#1{\vert#1\rangle}

\def\Longarrow{\protect\@lra}
\def\@lra{\relbar\joinrel\relbar\joinrel\relbar\joinrel%
          \relbar\joinrel\rightarrow}

\begin{document}

\title{Variational Quantum Non-Orthogonal Optimization}

\author{Pablo Bermejo}
\affiliation{Multiverse Computing, Paseo de Miram\'on 170, E-20014 San Sebasti\'an, Spain}
\affiliation{Donostia International Physics Center, Paseo Manuel de Lardizabal 4, E-20018 San Sebasti\'an, Spain}

\author{Rom\'an Or\'us}
\affiliation{Multiverse Computing, Paseo de Miram\'on 170, E-20014 San Sebasti\'an, Spain}
\affiliation{Donostia International Physics Center, Paseo Manuel de Lardizabal 4, E-20018 San Sebasti\'an, Spain}
\affiliation{Ikerbasque Foundation for Science, Maria Diaz de Haro 3, E-48013 Bilbao, Spain}
\affiliation{Corresponding author: roman.orus@dipc.org}

\begin{abstract} 

\bigskip
\centerline{{\bf Abstract}}
\bigskip 

Current universal quantum computers have a limited number of noisy qubits. Because of this, it is difficult to use them to solve large-scale complex optimization problems. In this paper we tackle this issue by proposing a quantum optimization scheme where discrete classical variables are encoded in non-orthogonal states of the quantum system. We develop the case of non-orthogonal qubit states, with individual qubits on the quantum computer handling more than one bit classical variable. Combining this idea with Variational Quantum Eigensolvers (VQE) and quantum state tomography, we show that it is possible to significantly reduce the number of qubits required by quantum hardware to solve complex optimization problems. We benchmark our algorithm by successfully optimizing a polynomial of degree 8 and 15 variables using only 15 qubits. Our proposal opens the path towards solving real-life useful optimization problems in today's limited quantum hardware. 

\end{abstract}

\maketitle

\emph{Introduction.-} Quantum computing is evolving rapidly thanks to significant advances in hardware. Current Noisy Intermediate-Scale Quantum (NISQ) processors \cite{nisq} are starting to show promise in specific tasks, including claims on quantum advantage \cite{qa1, qa2}, fault-tolerant gates \cite{faulttolerant}, and industrial applications for specific problems \cite{industrial}. Nonetheless, NISQ devices are still quite limited at the time of solving certain problems of relevance, such as complex optimization problems. This is true with the current implementation of quantum optimization algorithms such as Variational Quantum Eigensolvers (VQE) \cite{VQE} and Quantum Approximate Optimization Algorithms (QAOA) \cite{QAOA}. 

In this paper we attack this limitation by developing new quantum optimization schemes where classical variables are not encoded in orthogonal qubit states, but rather in other degrees of freedom of the quantum computer. An example is the case in which individual qubits are allowed to represent more than one classical bit variable each, by considering the degrees of freedom of the Bloch sphere \cite{latorre, clustering}. As we shall see, this allows to significantly reduce the number of qubits needed to solve complex discrete optimization problems in universal gate-based quantum computers, thus boosting the practical utility of NISQ devices for real-life problems. 

\bigskip

\emph{The problem.-} A key relevant problem for which NISQ devices are very limited is \emph{optimization}, which amounts to the minimization of a cost function, typically under some constraints which can also be included via, e.g.,  Lagrange multipliers. At the end of the day, an  arbitrary cost function $H$ of a discrete optimization problem takes the form
\beq
H \equiv f(q_0, q_2, \cdots q_{N-1}), 
\eeq
with $q_\alpha = 0, 1, \cdots, p-1$ a discrete variable that can take up to $p$ different values, with $\alpha = 0, 1, \cdots, N-1$. For example, if $p=2$ for all $\alpha$, then we have the case of an optimization problem with usual bit variables, $q_\alpha = 0, 1$ for all $\alpha$. Finding the minimum of such a cost function, for $N$ bit variables, and for a discrete polynomial of degree 2, is known as a Quadratic Unconstrained Binary Optimization (QUBO) problem and is well-known to be NP-Hard. Higher-order polynomials, and higher-dimensional variables, make the problem even harder. 

In many optimization schemes for NISQ devices, cost functions such as the one described above are typically considered by first boiling it down to bit variables. In a binary encoding scheme, this implies that the original variables $q_\alpha$ are expressed in terms of $m = \lceil \log_2(p) \rceil$ bits each, i.e., 
\beq
q_\alpha = \sum_{i = 0}^{m-1} 2^i x_{i,\alpha} ~~~ \forall \alpha.
\eeq
As an example, if $p = 4$, then we have that $m = 2$ and the correspondence $x_0 x_1 \rightarrow q$ between the individual bits and the original variable (for a given $\alpha$) is $[00 \rightarrow 0; 01 \rightarrow 1; 10 \rightarrow 2; 11 \rightarrow 3]$. With this decomposition in mind, the cost function $H$ can be written as 
\beq
H = f(x_{0,0}, x_{1,0}, \cdots, x_{m-1,N-1}), 
\eeq
in terms of $m \times N$ classical bit variables. 

As we can see, the number of bit variables quickly explodes in classical optimization problems. It is not surprising, therefore, that when solving such problems on quantum computers, their capabilities are quite limited if we consider one qubit in the quantum register for each bit variable of the cost function. Say, for instance, that you run a VQE algorithm on a universal gate-based quantum computer. The largest such machine built as of today is the IBM System One with 127 supercoducting qubits \cite{ibmq}. This implies that, with such an algorithmic approach, one can optimize cost functions up to 127 bits only, so that $m \times N = 127$. This is very far from real-life useful problems. As an example, a static portfolio optimization over the $N=500$  assets of the SP500 index, with $p = 8$ investment positions per asset would require of $m \times N = 1500$ qubits \cite{sp500}. 

\bigskip

\begin{figure}
  \centering
      \includegraphics[width=0.5\textwidth]{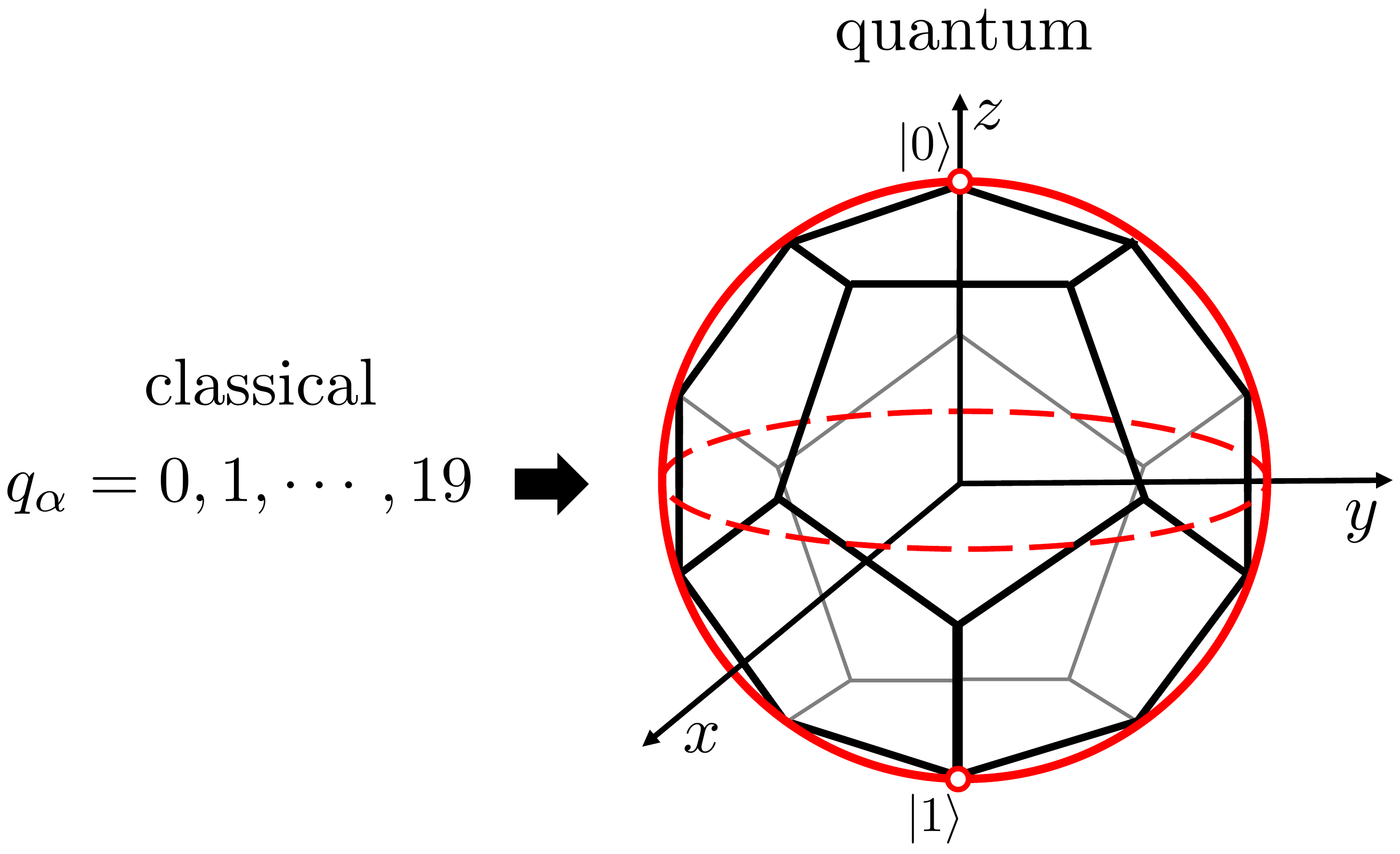}
  \caption{[Color online] The 20 configurations of classical variable $q_\alpha$ (left) are codified in 20 maximally-othogonal quantum states of a single qubit, which correspond to the 20 vertices of a dodecahedron inscribed in the qubit's Bloch sphere (right).} 
  \label{fig1}
\end{figure}

\emph{Method.-} Our proposal to overcome the above problem is to modify the assignment between the classical variables in the optimization problem and the quantum states in the quantum computer. As such, a quantum state of $N$ qubits has many more degrees of freedom than those in $N$ classical bit variables. If we were to assign one qubit per classical bit, then the correspondance of degrees of freedom would clearly be
\beq
\ket{0} \rightarrow 0, ~~~\ket{1} \rightarrow 1,  
\eeq
and this is a waste of quantum resources. However, this can be dramatically improved by \emph{assigning non-orthogonal quantum states to classical configurations} \cite{latorre}. The simplest example is the case of a single qubit: we can represent the different configurations of the classical discrete variable  $q_\alpha$ using different \emph{non-orthogonal} states of the qubit, i.e., by breaking the Bloch sphere into ``chunks". In particular, we can use $p$ { maximally-orthogonal states of one qubit to represent the values of the classical variable $q_\alpha = 0, 1, \cdots, p-1$ (such states are not mutually orthogonal, but minimize globally the mutual scalar products).} Such states  are maximally distinguishable by single-qubit measurements. These states also correspond to the vertices of convex polyhedra inscribed inside the qubit's Bloch sphere of the qubit \cite{latorre, clustering}, see Fig.\ref{fig1} for an example. In addition, this scheme can also be generalized to multi-qubit non-orthogonal quantum states. 

The readout of the classical configurations is thus done via quantum state tomography, which can be implemented in several ways. A simple approach is, again, to read out the quantum states of the individual qubits at each iteration of a variational quantum optimization algorithm, say VQE or QAOA. This would approximate the probability distribution $P(q_0, q_1,\cdots, q_{N-1})$, needed to estimate the expectation value of the cost function and its gradient at each iteration (epoch) of the variational algorithm, by    
\beq
P(q_0, q_1,\cdots, q_{N-1}) \approx P(q_0) P(q_1) \cdots P(q_{N-1}). 
\eeq
This approach, while missing correlations, is very efficient to implement and turns out to work quite well in many practical situations, at the expense of sometimes lowering the performance in the convergence of the variational quantum optimization algorithm. An alternative, more precise approach, is to implement quantum state tomography of the whole quantum state, which can be done using, e.g., compressed sensing \cite{compressedsensing}. 

{ The single-qubit measurements chosen here work well when the final state is not an entangled state, but rather a separable state. Forcing this type of measurements together with the optimization of the variational parameters in the quantum circuit, implies that the highest degree of entanglement typically happens towards the middle of the circuit evolution, with the qubits tending to be disentangling towards the end of the circuit. Notice also that, while multiqubit tomography could also be implemented, it is not efficient, in the sense that it cannot be scaled up in general for an arbitrary number of qubits.} 

\begin{figure}
  \centering
      \includegraphics[width=0.48\textwidth]{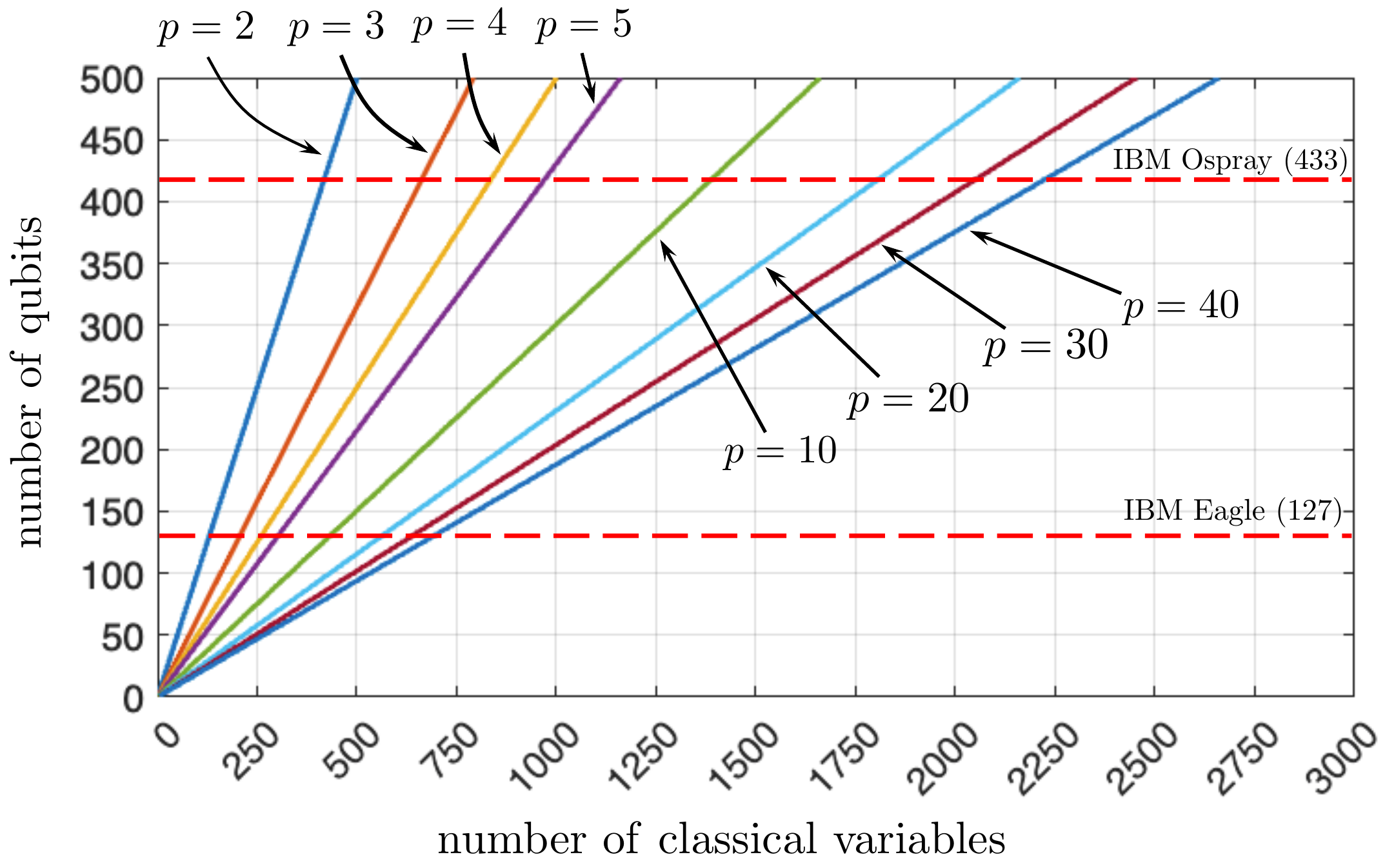}
  \caption{[Color online] Number of qubits required to represent a given number of classical binary variables, for different numbers $p$ of non-orthogonal quantum states per qubit. The dashed lines correspond to two IBM quantum processors: one with 127 qubits, which is already available, and one with 433 qubits, expected in principle before the end of 2022.} 
  \label{fig2}
\end{figure}

Using this simple idea, it turns out that we can fit much larger optimization problems in NISQ devices. As an example, take again the processor of 127 qubits from IBM, which is currently available. For this processor, and using 6 states per qubit, we could optimize cost functions of up to $\approx 328$ bit variables, being this already sufficient for a real-life static portfolio optimization problem of all the companies in the NASDAQ-100 index with 8 positions per asset. { To be more specific, we used the relation 
\begin{equation}
{\rm Number ~of ~bit ~variables} =  {\rm Number ~of ~qubits} \times \log_2 (p), 
\end{equation}
rounded to the closest integer from below.} In this scheme the technological problem is to be able to distinguish 6 different quantum states in the Bloch sphere of a single qubit. While this may sound harsh at first, notice that \emph{it is a single-qubit technical problem, and not a multi-qubit one.} As a matter of fact, distinguishing 6 states per qubit is also well within the capabilities of current quantum technology. And what is more: this can only improve as larger quantum processors are fabricated. For instance, for a 433 qubit quantum processor, such as the one planned for 2022 in the IBM quantum roadmap \cite{ibmq}, we could optimize the whole SP500 with 8 positions per asset \emph{with just 12 non-orthogonal states per qubit.} To understand this  graphically,  we show in Fig.\ref{fig2} the number of qubits required to represent a given number of classical binary variables, for different numbers $p$ of non-orthogonal quantum states per qubit.  

 \begin{figure}
  \centering
      \includegraphics[width=0.48\textwidth]{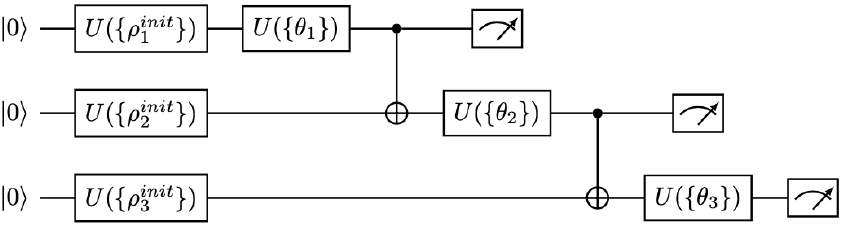}
  \caption{Example of variational quantum circuit used for the optimization of the 8-degree polynomial in Eq.(\ref{poly}), where the structure shown for 3 qubits should be generalized to 15 (we show only 3 for space reasons). This variational quantum circuit has a sequential, one-dimensional structure of entanglement, but more generic circuits can also be implemented. Measurements at the end of the circuit are meant to implement single-qubit quantum state tomography. The initial unitary operators set the initial quantum state, and the variational parameters $\{\theta_i \}$ correspond to single-qubit rotations for each qubit.} 
  \label{fig3}
\end{figure}

\bigskip

\begin{figure}
  \centering
      \includegraphics[width=0.48\textwidth]{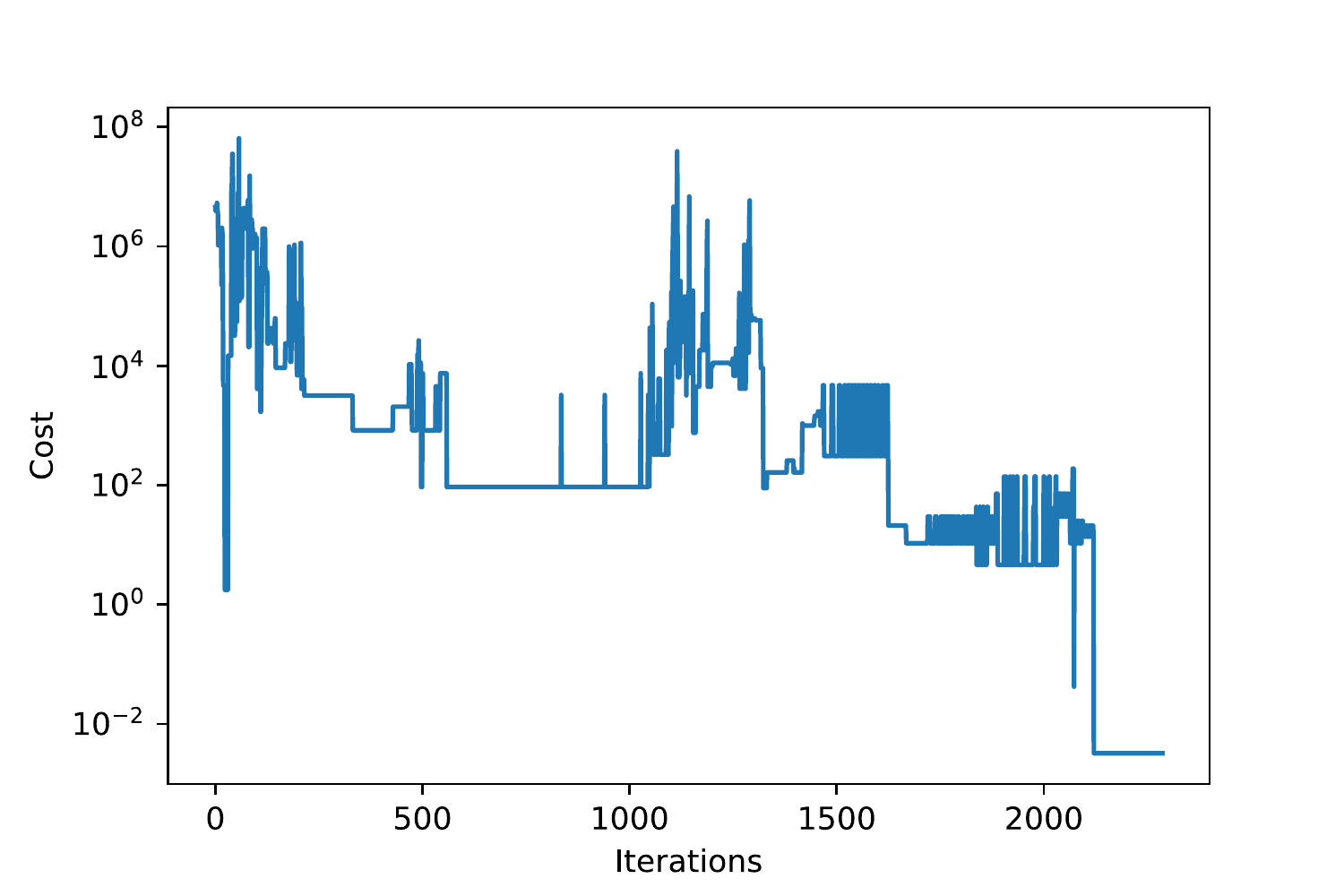}
  \caption{[Color online] Cost as a function of the number of iterations in the optimization of a degree-8 polynomial of 15 variables, with 20 possible values per variable, as explained in the text.} 
  \label{FigConv}
\end{figure}

\emph{Benchmark.-} We have benchmarked the idea by running a VQE optimization, on a classical simulator, to find the minimum of the following degree-8 polynomial: 
\beqa
H &=& (q_0^3 - q_1^3q_0 - q_2q_3 + q_4^3 - q_5q_6 - q_7^2 + q_8^3 \nonumber \\ 
&+& q_8q_{10}/5+ q_{11}q_1 + q_{12}^2 + q_{13}^3 + q_{14}q_0q_3)^2. 
\label{poly}
\eeqa
In the above equation, we consider the discrete values  
\beq
q_\alpha = -9, -8, \cdots, 10 ~~~ \forall \alpha, 
\eeq
so that each classical variable in the polynomial can take 20 different values. A standard VQE algorithm, codifying one bit per qubit, would need at least 65 qubits using a binary encoding of the variables, and all the bits in the cost function would be fully-connected, therefore implying deep and complex variational quantum circuits. Additionally, a quantum annealer would need thousands of qubits due to embedding for such a fully-connected graph. In our implementation, however, the simulation solves the minimization problem using 15 qubits and 20 quantum states per qubit with single-qubit tomography,  implying a reduction of 50 qubits with respect to VQE and thousands of qubits with respect to quantum annealing. Our simulation was able to find both the trivial and non-trivial minima of the cost function, converging with good accuracy after some epochs, and using a simple variational quantum circuit such as the one in Fig.\ref{fig3} but for 15 qubits. { The evolution of the cost as a function of the number of iterations is shown in Fig.\ref{FigConv}. The algorithm finds the optimal value in roughly $O(10^3)$ steps, which is to be compared to the $O(10^{20})$ steps that it would take for a bare classical sampling of $65$ bits, and the $O(10^{10})$ that it would take for an unstructured quantum search (so, even this simple implementation is seven orders of magnitude faster than a quantum search by using Grover's algortithm). The obtained minimum is $\approx 3 \times 10^{-3}$, with 6 circuit layers, and a learning rate of $0.008$ using Adam optimizer.}

In parallel to this example. we have also tested the performance of the algorithm for other problems, including non-polynomial cost functions and even with discontinuities, and the performance of the method has also been remarkably good. 

{ At this point, it is worth comparing the evolution of the cost in Fig.\ref{FigConv} to that of a standard VQE algorithm with $p=2$, and even with that of the continuum limit $p=\infty$, analyzed in detail in Ref.\cite{bermejo2022variational}. While for those two limiting cases one may observe a quite smooth behavior of the cost function, we do not find it for the intermadiate cases explored in this paper. We believe this is due to the fact that, in our current encoding, the $p$ maximally-orthogonal states of a qubit are distributed across the surface of Bloch's sphere, and are not naturally ordered. It should be possible to refine the embeddings and labelings of the states in the sphere, so as to improve the smoothness of the evolution. This will be explored in detail in further extensions of this work.} 

\bigskip 

\emph{Conclusions and outlook.-} In this paper we have presented a scheme to  significantly reduce the number of qubits needed in variational quantum optimization algorithms. This is based on the fact that configurations of the classical variables can be mapped to non-orthogonal quantum states of the quantum register, and these can be recovered via quantum state tomography. In the simplest case, we map several classical configurations to non-orthogonal states of each qubit, and perform single-qubit tomography. We have benchmarked this idea by successfully optimizing via VQE a polynomial function of 15 variables and 20 configurations per variable, using only 15 qubits. 

{ We stress that, similarly to other variational quantum algorithms, our method is heuristic. This means that there are only heuristic arguments about how this algorithm processes information, and the ultimate justification for using it, is that \emph{it works in practice}. And this is not a sloppy justification, since it is the same one being applied to all heuristic algorithms, even to neural networks and deep learning. Our algorithm here is no exception: it introduces an extra degree of freedom in the optimization (the number of states per qubit), that is indeed another parameter to play with in the heuristic algorithm. the benchmarks presented here, show that this heuristics works in practice, and may be useful in dealing with certain types of optimization problems. }

The scheme presented here is remarkably simple and powerful. As we have discussed throughout the paper, it dramatically reduces the number of qubits required to solve optimization problems in universal quantum computers. Traditionally, optimization problems had been better solved by means of quantum annealing approaches, with gate-based approaches being disregarded for optimization due to the low number of qubits. The fact that universal NISQ devices are getting more and more powerful (with 4000+ qubits being in the IBM roadmap for 2025) together with algorithmic improvements such as the one discussed here, is certainly an indication that universal quantum computers are closer to industrial applications than commonly expected. 4000 qubits with 40 non-orthogonal states per qubit could solve a classical  optimization problem of 200.000 bit variables, and these numbers are within reach in the near future for universal quantum computers. This is to the detriment of quantum annealers, in which this scheme cannot be applied as such and are therefore more limited in this respect. Additionally, optimization is the basis of many other algorithms, and the ideas discussed here are already being used as the basis of  further variational quantum algorithms solving a wide range of problems. This will be investigated in upcoming works. 

\bigskip

{\bf Acknowledgements.-} The authors acknowledge
DIPC, Ikerbasque, Basque Government and Diputaci\'on
de Gipuzkoa for constant support, as well as insightful discussions with the technical teams from Multiverse
Computing and DIPC on the algorithms and technical implementations. We also acknowledge Elena Zabala from Galbaian IP and Robert Harrison from Sonnenberg Harrison for supporting us with several patent applications in connection with parts of the work discussed here. 

\bigskip

{\bf Data availability.-} The datasets used and/or analysed during the current study are available from the corresponding author on reasonable request.

\bigskip
 
\onecolumngrid
\centering{{\bf References}}
\twocolumngrid
\bibliography{bibliography}

\end{document}